\documentclass[aps,onecolumn,superscriptaddress,nofootinbib]{revtex4-2}
\usepackage{amsmath,amssymb,graphicx,microtype,amsfonts,amstext,mathtools,physics,hyperref,bm}
\usepackage[normalem]{ulem}
\hypersetup{colorlinks=true,linkcolor=blue,citecolor=blue,urlcolor=blue}

\begin{document}
	
	\title{$\mathcal{H}$olographic $\mathcal{N}$aturalness and Pre-Geometric Gravity}
	
	\author{Andrea Addazi}
	\email{addazi@scu.edu.cn}
	\affiliation{Center for Theoretical Physics, College of Physics Science and Technology, Sichuan University, 610065 Chengdu, China}
	\affiliation{Laboratori Nazionali di Frascati INFN, Frascati (Roma), Italy, EU}
	
	\author{Salvatore Capozziello}
	\email{capozziello@na.infn.it}
	\affiliation{Dipartimento di Fisica  ``E.\ Pancini'', Università degli Studi di Napoli ``Federico II'', Compl.\ Univ.\ di Monte S.\ Angelo, Edificio G, Via Cinthia, I-80126, Napoli, Italy}
	\affiliation{Scuola Superiore Meridionale, Via Mezzocannone 4, I-80134, Napoli, Italy}
	\affiliation{Istituto Nazionale di Fisica Nucleare (INFN), Sez.\ di Napoli, Compl.\ Univ.\ di Monte S.\ Angelo, Edificio G, Via	Cinthia, I-80126, Napoli, Italy}
	
	\author{Giuseppe Meluccio}
	\email{giuseppe.meluccio-ssm@unina.it}
	\affiliation{Scuola Superiore Meridionale, Via Mezzocannone 4, I-80134, Napoli, Italy}
	\affiliation{Istituto Nazionale di Fisica Nucleare (INFN), Sez.\ di Napoli, Compl.\ Univ.\ di Monte S.\ Angelo, Edificio G, Via	Cinthia, I-80126, Napoli, Italy}
	
	\date{\today}
	
	\begin{abstract}
		The cosmological constant (CC, $\Lambda$) problem stands as one of the most profound puzzles in the theory of gravity, representing a remarkable discrepancy of about 120 orders of magnitude between the observed value of dark energy and its natural expectation from quantum field theory. This paper synthesizes two innovative paradigms holographic naturalness ($\mathcal{HN}$) and pre-geometric gravity (PGG)---to propose a unified and natural resolution to the problem. The $\mathcal{HN}$ framework posits that the stability of the CC is not a matter of radiative corrections but rather of quantum information and entropy. The large entropy $S_\textup{dS}\sim M_\textup{P}^2/\Lambda$ of the de Sitter (dS) vacuum (with $ M_\textup{P}$ being the Planck mass) acts as an entropic barrier, exponentially suppressing any quantum transitions that would otherwise destabilize the vacuum. This explains why the universe remains in a state with high entropy and relatively low CC. We then embed this principle within a pre-geometric theory of gravity, where the spacetime geometry and the Einstein--Hilbert action are not fundamental, but emerge dynamically from the spontaneous symmetry breaking of a larger gauge group, SO($1,4$)$\rightarrow$SO($1,3$), driven by a Higgs-like field $\phi^A$. In this mechanism, both $M_\textup{P}$ and $\Lambda$ are generated from more fundamental parameters. Crucially, we establish a direct correspondence between the vacuum expectation value (VEV) $v$ of the pre-geometric Higgs field and the de Sitter entropy: $S_\textup{dS}\sim v$ (or $v^3$). Thus, the field responsible for generating spacetime itself also encodes its information content. The smallness of $\Lambda$ is therefore a direct consequence of the largeness of the entropy $S_\textup{dS}$, which is itself a manifestation of a large Higgs VEV $v$. The CC is stable for the same reason a large-entropy state is stable: the decay of such state is exponentially suppressed. Our study shows that new semi-classical quantum gravity effects dynamically generate particles we call ``hairons'', whose mass is tied to the CC. These particles interact with Standard Model matter and can form a cold condensate. The instability of the dS space, driven by the time evolution of a quantum condensate, points at a dynamical origin for dark energy. This paper provides a comprehensive framework where the emergence of geometry, the hierarchy of scales and the quantum-information structure of spacetime are inextricably linked, thereby providing a novel and compelling path toward solving the CC problem.
	\end{abstract}
	
	\maketitle
	
	\section{Introduction}\label{sec1}
	The quest to understand the nature of dark energy, observed to be consistent with a small, positive cosmological constant (CC, $\Lambda$), has led to one of the most persistent challenges in fundamental physics: the CC problem. Quantum field theory suggests that the vacuum energy of the Standard Model (SM) contributes to the CC on scales as high as the Planck scale $M_\textup{P}$, resulting in a discrepancy with the observed value of approximately $10^{-120}$ \cite{Weinberg:1988cp}. Traditional approaches, which attempt to fine-tune away these radiative corrections, are widely regarded as unsatisfactory, especially in light of Weinberg's no-go theorem \cite{Weinberg:1988cp,Capozziello:2025bsm}.
	
	\textls[-15]{A transformative perspective emerges from the holographic principle and gravitational thermodynamics. The entropy of a de Sitter (dS) universe, given by \mbox{$S_\textup{dS} = 3\pi/G\Lambda \sim M_\textup{P}^2/\Lambda$,} with $G$ the Newtonian constant, is both finite and large (about $10^{120}$ for the observed value of the CC). This entropy counts the number of fundamental quantum degrees of freedom on the cosmological horizon. The holographic naturalness ($\mathcal{HN}$) paradigm \cite{Addazi:2020axm,Addazi:2020wnc,Addazi:2020mnm} leverages this insight by re-framing the problem. It argues that a transition from a state with a small CC (large  entropy $S_\textup{dS} \sim N \gg 1 $ with $N$ the number of qubits) to a state with a Planckian CC (unit entropy) is not just a loop correction but a quantum process that must overcome an exceptionally high entropic barrier. The amplitude for such a decay is exponentially suppressed, $\langle \Lambda | \Lambda_{\rm UV} \rangle \propto \exp(-S_\textup{dS}/2)$ (where $\Lambda_{\rm UV}$ is the CC of an ultraviolet (UV) completion of gravity and the angular brackets denote the scalar product of the Hilbert space), thus inherently protecting the small value of the  CC. From this viewpoint, one can assert that the standard computation of vacuum bubbles is incomplete, as it fails to account for the thermal, information-rich environment of the de Sitter horizon.}
	
	While $\mathcal{HN}$ explains the stability of the CC, it does not, by itself, explain its initial smallness. This is where the second pillar of our 
	framework, pre-geometric gravity (PGG), enters. Pioneered by Samuel 
	MacDowell and Fr\'ydoon Mansouri \cite{macdowell:unified} and Frank Wilczek \cite{wilczek:gauge}, PGG is based on the approach that the spacetime metric and Einstein's equations are not fundamental. Instead, they emerge from a more primitive, pre-geometric phase described by a gauge theory for a group like SO($1,4$) or SO($2,3$), devoid of a traditional metric. Formally, the dynamics of this phase is constructed using only the Levi-Civita symbol $\epsilon^{\mu\nu\rho\sigma}$, 
	which is a tensor density of weight $-1$ and the only intrinsically available object for defining covariant actions 
	\cite{Addazi:2024rzo,Addazi:2025vbw} (the Greek letter indices take the values 0 (for time) and  1, 2 or 3 (for spatial coordinates). See also \cite{akama:pregeometry, adler:einstein, bekenstein:gravitation, Chkareuli:2023krd, chamseddine:unification, diakonov:lattice, hehl:metric, 
		krasnov:spontaneous, Koivisto:2019ejt, Koivisto:2022uvd, Koivisto:2023epd, Koivisto:2024asr, maitiniyazi:irreversible, maitiniyazi:generation, Melichev:2023lwj, obukhov:diakonov, obukhov:dirac, verlinde:gravity, Westman:2013mf, Westman:2014yca, wetterich:spinors, wetterich:pregeometry, westman:cartan, Zlosnik:2018qvg} for other relevant studies and approaches on the subject.
	
	The key mechanism for PGG is spontaneous symmetry breaking (SSB). A scalar field $\phi^A$ acquires a vacuum expectation value (VEV) $\langle \phi^A \rangle = v \delta^A_4$, thereby breaking the gauge group down to the Lorentz group SO($1,3$). Through this Higgs mechanism, the components of the gauge field $A_\mu^{AB}$ of the initial group, either SO($1,4$) or SO($2,3$), are identified with the spin connection ($A_\mu^{ab}$) and the tetrads ($A_\mu^{a4}$); the capital Latin letter indices take the values $0$ to $4$ for internal-space coordinates in the unbroken phase, and the lower-case Latin letter indices take the values $0$ to $3$ for internal-space coordinates in the spontaneously broken phase. Remarkably, the effective actions, i.e., the MacDowell--Mansouri (MM) action or the Wilczek (W) action, yield the Einstein--Hilbert term, a cosmological constant term, and in one case (MM) also a Gauss--Bonnet term. The emergent Planck mass and the emergent CC are given,  
	respectively, as follows:
	
	\begin{itemize}
		\item[(i)] in the MM model: ~$\quad M_\textup{P}^2 \sim k_\textup{MM} v m^2$ \quad and ~$\quad \Lambda \sim m^2 \sim M_\textup{P}^2/k_\textup{MM} v$;
		\item[(ii)] in the W model: ~$\quad M_\textup{P}^2 \sim k_\textup{W} v^3 m^2$ \quad and $\quad \Lambda \sim m^2 \sim M_\textup{P}^2/k_\textup{W} v^3$.
	\end{itemize}
	Here, $k_\textup{MM}$ and $k_\textup{W}$ are the fundamental coupling constants of the PGG Lagrangians, while $m$ is a mass parameter introduced for defining the tetrad fields as dimensionless. Therefore, a larger VEV $v$ naturally leads to a smaller effective $\Lambda$, implementing a see-saw mechanism. Furthermore, a complete dictionary can be established to reconstruct the inverse metric $g^{\mu\nu}$ and the volume element $\sqrt{-g}$ from pre-geometric quantities \cite{Addazi:2024rzo}, which proves the full emergence of Riemannian geometry in the spontaneously broken phase.
	
	It is possible to perform a self-consistent and background-independent Hamiltonian analysis of theories of PGG \cite{Addazi:2025vbw}. Using the ADM formalism, it was demonstrated that the infrared limit of the spontaneously broken phase correctly reproduces all results of canonical general relativity. The constraint algebra was analyzed by employing Dirac's algorithm, which also allows us to count the degrees of freedom in the ultraviolet limit of the unbroken phase. This approach may lead to promising pathways for a UV completion of gravity, including a pre-geometric generalization of the Wheeler--DeWitt equation, a connection with Ashtekar variables, an extended BF theory and a generalized Plebanski gravity formulation. In addition, it was shown that, when the fundamental symmetry of spacetime is restored, 
	W-gravity may admit an ultraviolet fixed point where it becomes a topological theory, thus enabling its complete quantization \cite{Addazi:2025vbw}. The first exact solution to the pre-geometric field equations for a de Sitter-like universe was presented in Ref.\ \cite{Meluccio:2025uyo}, providing a potential resolution for the Big Bang singularity.
	
	The central synthesis presented in this paper is the identification of the pre-geometric Higgs VEV $v$ with the source of the de Sitter entropy: $S_\textup{dS}\sim v$ (or $v^3$). This creates a powerful synergy:
	
	\begin{itemize}
		\item[(i)] PGG explains the origin of the smallness of the CC via a see-saw mechanism requiring a correspondingly large $v$;
		\item[(ii)] $\mathcal{HN}$ explains the stability of such a small CC, as the large $v$ (and thus large $S_\textup{dS}$) imposes an exponential suppression $e^{-S_\textup{dS}} \sim e^{-|k_\textup{MM}|v}$ $\big(\text{or }\sim e^{-|k_\textup{W}|v^3}\big)$ on potentially `dangerous' transitions.
	\end{itemize}
	
	Here, we further develop this model by revealing that new pseudo-Nambu--Goldstone bosons, which we call ``hairons'' ($\varphi_h$), dynamically emerge from pre-geometric Wilson lines and instantons. These hairons, with mass $m_h\sim\sqrt\Lambda$, are identified as the fundamental quanta of information on the horizon. It is shown that, as dynamical fields, hairons provide a concrete realization of the quantum hairs of spacetime proposed by Gabriele Veneziano \cite{V1}, Sidney Coleman, John Preskill and Wilczek \cite{Coleman:1991ku}. Hairons interact with matter and gravity to thermalize vacuum fluctuations, providing an explanation for the naturalness of the CC. Furthermore, it is found that a condensate of hairons can not only account for the dS entropy but also, through its slow cosmological evolution, drive a dynamical form of dark energy.
	
	This paper is structured as follows. In Section \ref{sec2}, we review the $\mathcal{HN}$ argument for the stability of the CC. Section \ref{sec3} details the framework of PGG and the emergence of geometry through a mechanism of SSB. In Section \ref{sec4} we perform the crucial synthesis, linking the Higgs VEV $v$ to the de Sitter entropy $S_\textup{dS}$, as well as exploring the nature of hairons. Section \ref{sec5} confirms the consistency of the framework through a Hamiltonian analysis. Some remarks on $\mathcal{HN}$ phenomenology are then sketched out in Section \ref{sec6}. Finally, we discuss the main implications and draw our conclusions in Section \ref{sec7}.
	
	\section{$\mathcal{HN}$ and CC}\label{sec2}
	The holographic principle suggests that, in the case of the dS spacetime, the entropy is given by
	\begin{equation}
		\label{dS}
		S_\textup{dS}\sim \frac{A_H}{A_\textup{P}}\sim \frac{M_\textup{P}^{2}}{\Lambda}\sim N\, ,
	\end{equation}
	where $A_H$ is the area corresponding to the Hubble horizon and $N$ is the number of qubits of information holographically stored in spacetime. Equation \eqref{dS} suggests several essential insights.
	\begin{itemize}
		\item The amount of information in~vacuo is tremendously larger than that contained in baryonic matter and radiation. For instance, the CMB radiation entropy is approximately $10^{88}$, while Equation \eqref{dS} corresponds to a value of $10^{120}$ for the known value of the CC.
		\item From a thermodynamic perspective, $\sqrt\Lambda$ is interpreted as a temperature in vacuo, that is $T_\textup{dS}\sim \sqrt\Lambda \sim M_\textup{P}/\sqrt{N}$. Specifically, it is set by the Gibbons--Hawking temperature of the de Sitter horizon, namely $T_\textup{dS} = H/2\pi = \frac{1}{2\pi}\sqrt{\Lambda/{3}}$, where $H=\sqrt{\Lambda/{3}}$ is the Hubble parameter in the de Sitter space.
		\item The hierarchy between the CC and the Planck scale, $\Lambda/M_\textup{P}^2$, scales as the inverse of the spacetime entropy \eqref{dS}.
	\end{itemize}
	
	These points suggest a novel reinterpretation of the CC problem. Actually, a quantum transition from a state having a CC equal to the observed value $\Lambda$ to a state with a UV vacuum energy $\Lambda_{\rm UV}$, i.e.,\
	\begin{equation}
		\label{process}
		|\Lambda\rangle \rightarrow |\Lambda_{\rm UV}\rangle
	\end{equation}
	or 
	\begin{equation}
		\label{processA}
		|N\rangle \rightarrow |1\rangle,
	\end{equation}
	corresponds to a transition from a relatively large to a unit entropy, bearing in mind that $\Lambda_{\rm UV} \simeq M_\textup{P}^2$. Certainly, such a process is exponentially suppressed by the entropic barrier:
	\begin{equation}
		\label{entropicb}
		\langle \Lambda|\Lambda_{\rm UV}\rangle\sim e^{-S_\textup{dS}/2}\ll1\,.
	\end{equation}
	From Equation \eqref{dS}, Equation \eqref{entropicb} corresponds to 
	\begin{equation}
		\label{corr1}
		e^{-S_\textup{dS}}\sim e^{-N}\sim e^{-10^{120}}\, ,
	\end{equation}
	or, reinterpreted as a transition from $N\simeq 10^{120}$ to $N=1$, to
	\begin{equation}
		\label{int}
		\langle N|1\rangle \sim e^{-N/2}\simeq e^{-10^{120}/2}\,. 
	\end{equation}
	An amplitude \eqref{int} is associated with a quantum tunneling process which destroys $(N-1)$ qubits, so indeed the annihilation cost has to be an exponential suppression.
	
	Equations \eqref{entropicb} and \eqref{corr1} are consistent with the instantonic approach: the suppression, actually, corresponds to the quantum tunneling mechanism controlled by the gravitational Euclidean action, $S_{\rm E}$:
	\begin{equation}
		\label{EucliG}
		e^{-S_{\rm E}} \sim e^{-1/\alpha_G(\Lambda)}\, ,
	\end{equation}
	where $\alpha_{G}(\Lambda)=\Lambda/M_\textup{P}^{2}$ is the gravitational coupling constant at the energy scale $\sqrt{\Lambda}$. As is known, the $e^{-1/\alpha}$ instantonic suppression is universal; for instance, in Yang--Mills theories it is $e^{-1/\alpha_\textup{YM}}$, with the difference with gravity lying in the nature of the coupling and the renormalizability procedure. The gravitational interaction, in this sense, appears to be special since the inverse of the gravitational coupling constant $\alpha_{G}(\Lambda)$ scales as the dS entropy and, therefore, as the inverse of the number of qubits:
	\begin{equation}
		\label{qubb}
		\alpha_{G}(\Lambda)\sim S_\textup{dS}^{-1}\sim N^{-1}\,. 
	\end{equation}
	
	The quantum field theory computations of radiative corrections to the CC commonly found in the literature do not consider the entropy/quantum-information side of the CC problem. Thus, we reject them as incomplete and illusive. As discussed in Ref.\ \cite{Addazi:2020axm}, in order to perform bubble diagram computations including vacuum thermal effects, one can introduce new background fields sustaining quantum hairs, the so-called hairon field. Hairons `decorate' the de Sitter horizon with hairy qubits. The term ``hairon'' evokes the concept of quantum hairs on horizons \cite{V1,QH2,Coleman:1991ku}. While here we concretely realize this for the dS horizon, the same mechanism may be extendable to black hole horizons to address the information paradox---a promising direction for future investigations. The hairon fields are denoted by $\varphi_{h}({\bf x},t)$, with ${\bf x}$ being space coordinates on the dS boundary, and hairons possess an average thermal energy $\langle E_{h} \rangle= T_\textup{dS}\sim \sqrt{\Lambda}$. In the hairon dictionary, the dS spacetime then corresponds to a state of $N$ hairons, namely $|\varphi_{h_1},\dots,\varphi_{h_N}\rangle$. Actually, hairons can be thought of as a Bose--Einstein coherent condensate in case their self-interactions are negligible, with coherent fluctuations around the expectation value. It is worthy to note that such hairons are in a gapped phase with a degenerate ground state. In other words, hairons are in a state with a mass gap $\epsilon \sim \sqrt\Lambda/N$, while the expectation value $N \epsilon$ of such states is considerably large. The bubble diagrams of the SM receive the thermal insertions of hairons as background fields \cite{Addazi:2020axm,Addazi:2020wnc,Addazi:2020mnm}. Such an effective field theory approach renders comprehensible the CC stabilization from a quantum field theory prospective. Moreover, such an approach leads to cosmological implications too, like the metastability of dark energy that is discussed in Section \ref{sec3}. In the sections \ref{sec4}--\ref{sec6}, we then discuss how hairons naturally arise as modulus fields associated with orbifold gravitational instantons, rather than being particles introduced ad hoc.
	
	\section{Emergent Gravity in the Spontaneously Broken Phase}\label{sec3}
	
	We now consider a gauge theory formulated on a pre-geometric four-dimensional manifold, invariant under the de Sitter SO($1,4$) or the anti-de Sitter SO($2,3$) group. The fundamental dynamical variables are the components of the gauge potential $A_\mu^{AB}$ and the associated field strength $F_{\mu\nu}^{AB}$, which are antisymmetric in both their internal (Latin letter) and spacetime (Greek letter) indices. The objective is to dynamically generate a (pseudo-)Riemannian metric and the Einstein--Hilbert action without presupposing the existence of any spacetime metric or tetrads, while also strictly adhering to the principle of general covariance. The only fixed structure is an internal-space metric $\eta$ with signature $(-,+,+,+,\pm)$ (depending on the gauge group), which generalizes the Minkowski metric in five dimensions. This emergence is achieved via the Higgs mechanism, employing a spacetime scalar field $\phi^A$ that transforms as a vector in the internal space. The SSB of its 
	ground state, actually, can reduce the original gauge symmetry to that of the Lorentz group SO($1,3$), thereby recovering the Einstein equivalence principle in the spontaneously broken phase \cite{wilczek:gauge}. In this geometric phase, the curvature of spacetime and its dynamics effectively emerge from the interactions of the pre-geometric fields $A_\mu^{AB}$ and $\phi^A$ below a specific energy scale. Crucially, SO($1,4$) (or SO($2,3$)) is an internal gauge group on a four-dimensional manifold, not a spacetime symmetry. Thus, the symmetry-breaking pattern SO($1,4$)$\rightarrow$SO($1,3$) is fundamentally different from Kaluza--Klein compactifications, as it generates geometry itself rather than deriving from some higher-dimensional geometry.
	
	\subsection{Pre-Geometric Actions}
	To construct generally covariant Lagrangian densities, one must form scalar densities of weight $-1$. Since the fundamental fields $A_\mu^{AB}$ and $F_{\mu\nu}^{AB}$ are covariant tensors, contracting their indices requires a contravariant object. In the absence of an inverse metric, the only intrinsically defined four-dimensional contravariant object on the manifold is the constant Levi-Civita symbol $\epsilon^{\mu\nu\rho\sigma}$, a tensor density of weight $-1$. The term `intrinsically' refers to the feature that no structure other than the differential structure of the manifold is necessary---in particular, a metric structure is not required. The density character of actions is defined solely through the Jacobian determinant of coordinate transformations, with no reference to the spacetime metric. Consequently, the Levi-Civita symbol, used to its first power, is the fundamental object for constructing generally covariant Lagrangian densities, as it contracts exactly four covariant indices and possesses the correct weight. For any pre-geometric Lagrangian density, the corresponding action is then obtained by integrating over all the points of the four-dimensional spacetime manifold.
	
	\textls[-25]{Using the Levi-Civita symbol, two distinct pre-geometric Lagrangian densities can be formulated for the unbroken phase. The first one, introduced by MacDowell and Mansouri \cite{macdowell:unified}, is}
	\begin{equation}
		{L}_\textup{MM} = k_\textup{MM} \epsilon_{ABCDE} \epsilon^{\mu\nu\rho\sigma} F_{\mu\nu}^{AB} F_{\rho\sigma}^{CD} \phi^E,
	\end{equation}
	while the second one, proposed by Wilczek \cite{wilczek:gauge}, is
	\begin{equation}
		{L}_\textup{W} = k_\textup{W} \epsilon_{ABCDE} \epsilon^{\mu\nu\rho\sigma} F_{\mu\nu}^{AB} \nabla_\rho \phi^C \nabla_\sigma \phi^D \phi^E.
	\end{equation}
	Here, $\nabla_\mu$ denotes the gauge covariant derivative, acting on internal-space vectors as
	\begin{equation}\label{eq:cov_dev-new}
		\nabla_\mu \phi^A = \partial_\mu \phi^A + A_{B\mu}^A \phi^B = (\delta^A_B \partial_\mu + A^A_{B\mu}) \phi^B,
	\end{equation}
	where $A_{B\mu}^A \equiv \eta_{BC} A_\mu^{AC}$; here, $\partial_\mu \equiv \partial / \partial x_\mu$ and $\delta^A_B$ is the Kronecker delta. The mass dimensions of the coupling constants are $[k_\textup{MM}] = [\phi]^{-1}$ and $[k_\textup{W}] = [\phi]^{-3}$, with the dimensions of the field $\phi^A$ initially left unspecified. Before analyzing the SSB mechanism itself, let us first examine the effective theory emerging after the symmetry reduction from SO($1,4$) or SO($2,3$) to SO($1,3$). This allows for a straightforward understanding of the classical physics implications of these theories, before venturing into the pre-geometric regime.
	
	The SSB mechanism selects a preferential direction in the internal space, characterized by a fixed VEV $\langle \phi^A \rangle= v \delta^A_4$, where the index $4$ is fixed and $v$ is a nonzero constant. This breaking allows for a classification of the components of the gauge potential: for each spacetime index $\mu$, four potentials are of the type $A_\mu^{A4} \equiv A_\mu^{a4}$ and the remaining six are $A_\mu^{AB} \equiv A_\mu^{ab}$ (with $A, B \ne 4$). Let us first compute the form of ${L}_\textup{MM}$ after the SSB. Upon making the identifications
	\begin{equation}\label{eq:identifications-new}
		e_\mu^a \equiv m^{-1} A_\mu^{a4}, \qquad \omega_\mu^{ab} \equiv A_\mu^{ab},
	\end{equation}
	where $e_\mu^a$ and $\omega_\mu^{ab}$ are the tetrads and the spin connections, respectively, and a mass parameter $m$ is introduced for dimensional consistency, the Lagrangian density decomposes into three distinct terms:
	\begin{equation}
		\label{Lmm}
		{L}_\textup{MM} \xrightarrow{\textup{SSB}} \pm 16k_\textup{MM} v m^2 e e_a^\mu e_b^\nu R_{\mu\nu}^{ab} - 96k_\textup{MM} v m^4 e - 4k_\textup{MM} 
		v e \mathcal{G}
		,
	\end{equation}
	where `$\pm$' distinguishes respectively between the groups SO($1,4$) or SO($2,3$), $e$ represents the tetrad determinant and $\mathcal{G}$ is the Newtonian constant. The terms in the Lagrangian \eqref{Lmm} correspond to the Einstein--Hilbert action, the cosmological constant term and the Gauss--Bonnet term, respectively. For consistency with conventional consideration, the Planck mass must then be identified as
	\begin{equation}
		\label{Mplmm}
		M_\textup{P}^2 \equiv \pm 32 k_\textup{MM} v m^2.
	\end{equation}
	The expression \eqref{Mplmm} indicates the emergent nature of the Planck scale, as it arises from a specific combination of the fundamental parameters $k_\textup{MM}$, $v$ and $m$. Consequently, the CC is also emergent, given by
	\begin{equation}
		\label{Lmbdmm}
		\Lambda \equiv \pm 3m^2 = \frac{3M_\textup{P}^2}{32 k_\textup{MM} v}.
	\end{equation}
	The expression \eqref{Lmbdmm} reveals a natural see-saw suppression mechanism. Assuming the experimentally measured value $M_\textup{P}^2 \sim 10^{37}$ GeV$^2$ and a coupling constant of order unity ($k_\textup{MM} \sim \pm 1\,[\phi]^{-1}$), the observed small value $\Lambda \sim 10^{-84}$ GeV$^2$ can be generated from a large vacuum expectation value $v \sim 10^{119}\,[\phi]$. Within this framework, the CC is therefore set by the mass scale $m$ of the SSB.
	
	In the Wilczek model, the analysis of the SSB for ${L}_\textup{W}$ follows a similar way as that for ${L}_\textup{MM}$. One additional element is required: the action of the covariant derivative on the internal-space vector $\phi^A$ after its VEV becomes fixed. From Equation \eqref{eq:cov_dev-new}, one finds
	\begin{equation}\label{eq:cov-der-new}
		\nabla_\mu \phi^A \xrightarrow{\textup{SSB}} v \nabla_\mu \delta^A_4 = A_{B\mu}^A \delta^B_4= v A_{4\mu}^A = \pm v A_\mu^{a4}.
	\end{equation}
	Utilizing the identifications \eqref{eq:identifications-new} and following a computation similar to that for the MM model, one arrives at the following result:
	\begin{equation}
		{L}_\textup{W} \xrightarrow{\textup{SSB}} -4k_\textup{W} v^3 m^2 e e^\mu_a e^\nu_b R_{\mu\nu}^{ab} \pm 48k_\textup{W} v^3 m^4 e.
	\end{equation}
	This theory yields precisely the Einstein--Hilbert Lagrangian density plus the cosmological constant term, with no Gauss--Bonnet contribution. The Planck mass and the CC are then identified as
	\begin{equation}
		M_\textup{P}^2 \equiv -8 k_\textup{W} v^3 m^2 \qquad \text{and} \qquad \Lambda \equiv \pm 6m^2 = \mp \frac{3M_\textup{P}^2}{4 k_\textup{W} v^3}.
	\end{equation}
	\textls[-25]{As for Equation \eqref{Lmbdmm}, assuming a coupling of order unity (\mbox{$k_\textup{W} \sim -1\,[\phi]^{-3}$}), the observed value \mbox{$\Lambda \sim 10^{-84}$ GeV$^2$} emerges from a large VEV $v \sim 10^{40}\,[\phi]$.}
	
	\subsection{Geometric and Pre-Geometric Variables}
	To demonstrate that a metric theory of gravity can emerge, completely and consistently, from a non-metric field theory, it is essential to derive within the same formalism not only an effective metric $g_{\mu\nu}$, but also its inverse $g^{\mu\nu}$ and the tetrad determinant $e=\sqrt{-g}$, with $g$ being the determinant of the metric tensor. The inverse metric is indispensable for constructing kinetic and interaction terms in the Lagrangians of general relativity and the SM, while the metric determinant is required for a generally covariant volume element in the action principle. The metric can be constructed from quadratic combinations of tetrads. Similarly, the inverse metric arises from quadratic combinations of inverse tetrads.
	
	The definition of an emergent tetrad was provided in the identifications \eqref{eq:identifications-new}, and subsequently allows for the construction of an effective metric:
	\begin{equation}
		m^{-2}\eta_{AB}A_\mu^{A4}A_\nu^{B4} \xrightarrow{\textup{SSB}} \eta_{ab}e_\mu^a e_\nu^b \equiv g_{\mu\nu}.
	\end{equation}
	In light of the expression \eqref{eq:cov-der-new}, an effective metric can also be formulated as
	\begin{equation}
		v^{-2}m^{-2}P_{\mu\nu} \xrightarrow{\textup{SSB}} \eta_{ab}e_\mu^a e_\nu^b \equiv g_{\mu\nu},
	\end{equation}
	where the auxiliary field $P_{\mu\nu}$ is defined as
	\begin{equation}\label{eq:aux_P}
		P_{\mu\nu} \equiv \eta_{AB} \nabla_\mu \phi^A \nabla_\nu \phi^B.
	\end{equation}
	
	To obtain a suitable expression for an inverse tetrad, let us first define the auxiliary field
	\begin{equation}\label{eq:aux_w}
		w_A^\mu \equiv \pm \epsilon_{ABCDE} \epsilon^{\mu\nu\rho\sigma} \nabla_\nu \phi^B \nabla_\rho \phi^C \nabla_\sigma \phi^D \phi^E.
	\end{equation}
	After the SSB, the field \eqref{eq:aux_w} becomes
	\begin{equation}
		w_A^\mu \xrightarrow{\textup{SSB}} v^4 \epsilon_{ABCDE} \epsilon^{\mu\nu\rho\sigma} \delta_4^E A_\nu^{B4} A_\rho^{C4} A_\sigma^{D4} = v^4 m^3 \epsilon_{abcd} \epsilon^{\mu\nu\rho\sigma} e_\nu^b e_\rho^c e_\sigma^d,
	\end{equation}
	which is effectively proportional to the inverse tetrad multiplied by the tetrad determinant. This can be verified by considering the identity $e_a^\mu e_\lambda^a = \delta_\lambda^\mu$:
	\begin{equation}\label{eq:w}
		w_a^\mu e_\lambda^a \xrightarrow{\textup{SSB}} v^4 m^3 \epsilon_{abcd} \epsilon^{\mu\nu\rho\sigma} e_\lambda^a e_\nu^b e_\rho^c e_\sigma^d = -6 v^4 m^3 e \delta_\lambda^\mu.
	\end{equation}
	The result \eqref{eq:w} might suggest that the inverse tetrad and the tetrad determinant are not independent. However, an effective tetrad determinant can be constructed independently via another auxiliary field:
	\begin{equation}\label{eq:aux_J}
		J \equiv \epsilon_{ABCDE} \epsilon^{\mu\nu\rho\sigma} \nabla_\mu \phi^A \nabla_\nu \phi^B \nabla_\rho \phi^C \nabla_\sigma \phi^D \phi^E,
	\end{equation}
	which can be interpreted as an effective Jacobian determinant. After the SSB, one finds
	\begin{equation}\label{eq:J}
		J \xrightarrow{\textup{SSB}} v^5 \epsilon_{ABCDE} \epsilon^{\mu\nu\rho\sigma} \delta_4^E A_\mu^{A4} A_\nu^{B4} A_\rho^{C4} A_\sigma^{D4} = v^5 m^4 \epsilon_{abcd} \epsilon^{\mu\nu\rho\sigma} e_\mu^a e_\nu^b e_\rho^c e_\sigma^d = -24 v^5 m^4 e.
	\end{equation}
	Finally, an effective inverse metric can be constructed as follows:
	\begin{equation}\label{eq:inverse_metric}
		16 v^2 m^2 J^{-2} \eta^{AB} w_A^\mu w_B^\nu \xrightarrow{\textup{SSB}} \eta^{ab} e_a^\mu e_b^\nu \equiv g^{\mu\nu}.
	\end{equation}
	
	
	To summarize, a comprehensive dictionary is presented in this section to translate the pre-geometric fields $A_\mu^{AB}$ and $\phi^A$ into the effective geometric fields $e_\mu^a$ and $g_{\mu\nu}$ after the SSB has been established. For conciseness, let us reiterate the complete dictionary here:
	\begin{equation}\label{eq:dictionary}
		\begin{split}
			m^{-1} A_\mu^{A4} \quad \text{or} \quad \pm v^{-1} m^{-1} \nabla_\mu \phi^A \quad &\xrightarrow{\textup{SSB}} \quad e_\mu^a, \\
			4 v m J^{-1} w_A^\mu \quad &\xrightarrow{\textup{SSB}} \quad e_a^\mu, \\
			-\frac{1}{24} v^{-5} m^{-4} J \quad &\xrightarrow{\textup{SSB}} \quad e = \sqrt{-g}, \\
			m^{-2} \eta_{AB} A_\mu^{A4} A_\nu^{B4} \quad \text{or} \quad v^{-2} m^{-2} P_{\mu\nu} \quad &\xrightarrow{\textup{SSB}} \quad \eta_{ab} e_\mu^a e_\nu^b \equiv g_{\mu\nu}, \\
			16 v^2 m^2 J^{-2} \eta^{AB} w_A^\mu w_B^\nu \quad &\xrightarrow{\textup{SSB}} \quad \eta^{ab} e_a^\mu e_b^\nu \equiv g^{\mu\nu},
		\end{split}
	\end{equation}
	with the auxiliary fields $P_{\mu\nu}$, $w_A^\mu$ and $J$ defined in Equations \eqref{eq:aux_P}, \eqref{eq:aux_w} and \eqref{eq:aux_J}, respectively. It is straightforward to verify that all the emergent geometric quantities transform as tensors and that the effective metric and its inverse are symmetric.
	
	\subsection{The Symmetry-Breaking Potential}
	The process of SSB, which reduces the gauge symmetry from SO($1,4$) or SO($2,3$) to SO($1,3$) via the field $\phi^A$, is responsible for the dynamical emergence of a classical spacetime metric in theories of PGG. This mechanism can be implemented by introducing a symmetry-breaking potential term into the Lagrangian density:
	\begin{equation}\label{eq:potential}
		{L}_\textup{SB} = -k_\textup{SB} v^{-4} \lvert J\rvert (\eta_{AB}\phi^A\phi^B \mp v^2)^2,
	\end{equation}
	where $k_\textup{SB}$ is a positive constant with dimension $[k_\textup{SB}] = [\phi]^{-5}$. The potential $-\mathcal{L}_\textup{SB}$ is minimized, and the term \eqref{eq:potential} is stationarized, for field configurations satisfying $\eta_{AB}\phi^A\phi^B = \pm v^2$. A specific solution, such as $\phi^A = v\delta_4^A$, can be chosen; then, any other VEV related to this one by a gauge transformation is physically equivalent \cite{wilczek:gauge}. The factor of $\lvert J\rvert$ ensures that $\mathcal{L}_\textup{SB}$ transforms as a scalar density, thus preserving general covariance. It is noteworthy that if one imposes a unimodular condition, as performed by Wilczek \cite{wilczek:gauge}, the $\lvert J\rvert$ factor can be omitted. In that case, the coupling constant $k_\textup{SB}$ assumes a fixed mass dimension of $[M]^4$, independently of the chosen dimensions for the field $\phi^A$. The field $\phi^A$ is quantized by expanding it around its VEV as $\phi^A = (v + \rho)\delta_4^A$, which defines the unitary gauge. In this gauge, the four to-be Goldstone bosons associated with the broken generators can be absorbed via the Higgs mechanism, leaving a single scalar degree of freedom $\rho$.
	
	An alternative mechanism for achieving the SSB, which circumvents the introduction of an explicit potential, was explored in Ref.\ \cite{Addazi:2025qkc}. This approach posits that the field $\phi^A$ can dynamically evolve toward a fixed VEV through a gradient descent process. This relaxation is mathematically governed by a set of Langevin equations, situating the mechanism within the broader context of stochastic quantization.
	
	\section{$\mathcal{HN}$ and PGG}\label{sec4}
	Under the perspective adopted in this study, it is an intriguing possibility to connect $\mathcal{HN}$ with a theory of PGG \cite{Addazi:2024rzo}. Actually, the two proposals appear to be complementary. On one hand, $\mathcal{HN}$ cannot explain the smallness of the CC but, assuming it, can explain its stabilization, while on the other hand, PGG can explain the emergence of a positive CC via a symmetry-breaking pattern SO($1,4$)$\rightarrow$SO($1,3$).
	
	As remarked in Sections \ref{sec1} and \ref{sec3}, in both the MM and the W models of PGG, the Planck scale and the CC are emergent, with different parametrizations:
	\begin{gather}
		\text{MM model:} \quad M_\textup{P}^2 \equiv 32 k_\textup{MM} v m^2 \quad \text{and} \quad \Lambda \equiv 3m^2 = \frac{3M_\textup{P}^2}{32 k_\textup{MM} v};\\
		\text{W model:}\quad M_\textup{P}^2 \equiv -8 k_\textup{W} v^3 m^2 \quad \text{and} \quad \Lambda \equiv 6m^2 = -\frac{3M_\textup{P}^2}{4 k_\textup{W} v^3}.
	\end{gather}
	Interestingly, in the MM model, the VEV $v$ is inversely proportional to the gravitational coupling at an energy scale corresponding to the CC; in W-gravity, instead, it is the cubic power which is involved in the same relation, namely:
	\begin{gather}
		\text{MM model:}\quad\alpha(\Lambda)\sim \frac{1}{|k_\textup{MM}|v} \sim 10^{-120},\\
		\text{W model:}\quad \alpha(\Lambda)\sim \frac{1}{|k_\textup{W}|v^{3}} \sim 10^{-120}\, .
	\end{gather}
	However, as seen in Section \ref{sec2}, the entropy of a dS universe is the inverse of the gravitational coupling \eqref{qubb}). Therefore, the dS entropy can be suggestively related to the new VEV scale of PGG: \vspace{-3pt}
	\begin{gather}
		\text{MM model:}\quad S_\textup{dS}\sim |k_\textup{MM}|v \sim 10^{120},\\
		\text{W model:}\quad S_\textup{dS}\sim |k_\textup{W}|v^{3} \sim 10^{120}\, .
	\end{gather}
	Note that, if the field $\phi^{A}$ has a non-canonical mass dimension $[\phi]=[M]^0$ as proposed by Wilczek \cite{wilczek:gauge}, then the VEV is just a considerably large number. It may be significant to observe that the information contained in the dS vacuum here acquires a pre-geometric meaning as the VEV of the new Higgs-like field. Therefore, the SSB mechanism of PGG generates not only the spacetime metric, its relativistic dynamics, the fundamental gravitational energy scales (the Planck mass and the CC) and diffeomorphism invariance, but also the dS entropy.
	
	This result suggests that the new $\phi^A$ field can serve as an information field, or in other words, it sustains the spacetime qubits when it is in this spontaneously broken phase. In this sense, the hairon field, which is related to the square root of the entropy, may be reinterpreted as a mean field originating from the pre-geometric Higgs-like field. Moreover, as a consequence of Equation \eqref{corr1}, the combined framework of $\mathcal{HN}$ plus PGG entails that the comparably small value of the CC is protected since any destabilization due to quantum effects is exponentially suppressed as
	\vspace{-3pt}
	\begin{gather}
		\text{MM model:}\quad e^{-S_\textup{dS}}\sim e^{-|k_\textup{MM}|v},\\
		\text{W model:}\quad e^{-S_\textup{dS}}\sim e^{-|k_\textup{W}|v^3}\, .
	\end{gather}
	The pre-geometric phase ($\langle \phi^4 \rangle = 0$) is a state of no geometry and minimal entropy, i.e., no quantum information. The Higgs phase ($\langle \phi^4 \rangle = v \neq 0$) is a state that has spontaneously generated both the spacetime geometry and a quite large amount of entropy/quantum information, as $S_\textup{dS} \sim v$ in the MM model (or $v^3$ in the W model). This makes the smallness of $\Lambda$ a consequence of the vastness of the quantum-information content of the universe.
	

	
	
	Let us now scrutinize and better understand the correspondence between the VEV of $\phi^4$ and the dS entropy from multiple perspectives. The emergence of $N$ qubits of information has been understood in light of a discretization of the dS boundary into Planckian-sized pixels, as a consequence of the emergence of the Planck length itself after the SSB. Nevertheless, so far it is not quite understandable if fields like hairons can also naturally emerge within the context of a theory of PGG.
	
	One possibility is that hairons are related to moduli of non-perturbative effects such as gravitational instantons in orbifolds. We argue that a $\mathbb{Z}_N$ symmetry, with $N$ related to the dS entropy, can be obtained from the evaluation of Wilson loops wrapping around a dS instanton. It is known that the dS spacetime corresponds to a gravitational instanton solution which is just the Euclidean dS with an $S_{4}$ topology (hypersphere-equivalent) of radius $R \sim 1/\sqrt\Lambda$ \cite{EdS}. This solution is obtained after a Wick rotation for time, $t\rightarrow i\tau$, and the period of this coordinate is found to be $\beta=1/T_\textup{dS}\sim 1/\sqrt\Lambda$, i.e., the inverse of the Gibbons--Hawking temperature \cite{EdS1,EdS2} (see also \cite{Bousso:1998bn,Bousso:1999ms,Bousso:2000nf}).
	
	In gauge theories, the Wilson loop expectation value in the presence of an instanton is expressed as a path integral:
	\begin{equation}
		\label{WLL}
		\langle W[\gamma] \rangle = \frac{\int \mathcal{D}A \, W[\gamma] \, e^{-S[A]} \delta(F[A] - \tilde{F}[A])}{\int 
			\mathcal{D}A \, e^{-S[A]} \delta(F[A] - \tilde{F}[A])}.
	\end{equation}
	Here, $S[A]$ is the Yang--Mills action and the $\delta(F[A] - \tilde{F}[A])$ term enforces the self-duality condition $F_{\mu\nu} = \tilde{F}_{\mu\nu}$ for the instanton configuration, with $\delta(a-b)$ being the Dirac delta function. For an instanton of charge $Q=K/N$, with $K$ being the topological number and $N$ the rank of the gauge group, the classical action $S[A]$ is proportional to $Q$,
	\begin{equation}
		\label{InstantonAct}
		S[A] = \frac{8\pi^2}{g^2} \frac{K}{N}\sim \frac{1}{\alpha_\textup{YM}}\frac{K}{N},
	\end{equation}
	as obtained from imposing the self-duality constraint and using the definition of the topological number $K$. For relatively large Wilson loops, the phase \eqref{WLL} is given by
	\begin{equation}
		\label{lWg}
		\langle W[\gamma] \rangle \sim \exp\left( -2\pi i \frac{K}{N} \right).
	\end{equation}
	The factor $\exp(-2\pi i K/N)$ arises because the instanton modifies the holonomy of the gauge field around the loop, introducing a phase due to the non-trivial topological charge. The solutions are known in the literature as fractional gauge instantons \cite{FGI1,FGI2,tHooft:1981nnx,FGI3,FGI4,FGI5,FGI6,FGI7,FGI10,FGI11,FGI12,FGI13} and are typically obtained on manifolds with twisted boundary conditions. The residual central symmetry of SU($N$) is, in this case, $\mathbb{Z}_N$.
	
	This consideration can suggest the existence of a new gravitational instanton, on top of the standard Euclidean dS instanton, which can be obtained as the orbifold of the hypersphere O\textsubscript{4,$N$}$=$$S_4/\mathbb{Z}_N$. O\textsubscript{4,$N$} is a hypersphere with $N$ conical singularities on the surface, which in turn correspond topologically to a set of two-surface $\Sigma$. This new class of instantons can be related to the known asymptotically locally Euclidean (ALE) instantons \cite{ALE1,ALE2,ALE3,ALE4,ALE5}. The connection between $S^4/\mathbb{Z}_N$ orbifolds and ALE instantons emerges through a local analysis in the vicinity of the orbifold singularities. While $S^4/\mathbb{Z}_N$ constitutes a compact orbifold, the geometry in the immediate vicinity of each singularity is locally modeled by $\mathbb{R}^4/\mathbb{Z}_N$. Instantons on $S^4/\mathbb{Z}_N$, when lifted to the covering space $S^4$ and projected to $\mathbb{R}^4$ via a stereographic projection, yield $\mathbb{Z}_N$-symmetric instanton configurations on $\mathbb{R}^4$. The local behavior in the vicinity of a fixed point precisely defines an orbifold instanton on $\mathbb{R}^4/\mathbb{Z}_N$, which in turn represents the singular limit of a smooth, hyper-K\"ahler ALE instanton upon resolution of the singularity. Consequently, the global $S^4/\mathbb{Z}_N$ instanton solution can be interpreted as a compact framework that encapsulates the local data of $A_{N-1}$-type ALE instantons situated at singular points of this framework, with the fractional instanton numbers being characteristic of the orbifold that is accounted for by the Kronheimer--Nakajima construction upon desingularization \cite{ALE1,ALE4,ALE5}.
	
	
	The main difference between $S_4$ instanton spacetime and $S_4/\mathbb{Z}_N$ is in the modulus space. $S_4$ instantons have a modulus space trivially corresponding to the center of the instantons, i.e., $\text{dim}(S_4)=4$, which in turn corresponds to the dimension of the coset SO($5$)/SO($4$), with SO($5$) being the group of $S_4$ isometries. This does not match the number of moduli for $S_4/\mathbb{Z}_N$, which scales as $N$. More precisely, $S_4/\mathbb{Z}_N$ has the same modulus space dimension of $\mathbb{Z}_N$ ALE instantons, which corresponds to ${M}(S_4,N) \sim 3N$ in the case the instanton winding number being $K=1$, counting the position of the fixed orbifold singularities. Indeed, there is a corresponding number of Nambu--Goldstone bosons related to the spontaneous symmetry breaking of the isometries, scaling as $N$.
	
	Let us note that $S_4/\mathbb{Z}_N$ is equivalent to the geometry of $N$ conic singularities with
	\begin{equation}
		\label{Thetaaa}
		\Theta=2\pi( 1- \Psi),\qquad \Psi=8\pi G\mu\sim \frac{1}{N}\, , 
	\end{equation}
	where $\Theta$ and $\Psi$ are the deficit and opening angles, respectively, while $\mu$ is the Euclidean string tension which sustains the conic singular geometry. In the case of the dS geometry, Euclidean strings have a tension which is proportional to $\Lambda$, $\mu \sim \Lambda$. In particular, in the limit of $N\rightarrow \infty$, the Euclidean dS geometry is exactly obtained. Indeed, $\Theta_\textup{dS}=2\pi$ reflects the periodicity of the dS time in Euclidean spacetime. Let us also notice that Equation \eqref{Thetaaa} implies that
	\begin{equation}
		\label{implyy}
		\Psi S \sim 1\, , 
	\end{equation}
	which can be interpreted as an uncertainty relation between the deficit angle and the entropy or between the angle and the number $N$ of qubits.
	
	The metric for the $(3+1)$-dimensional Euclidean de Sitter space is
	\begin{equation}
		\label{EdS}
		ds_{\rm E}^2=\bigg(1-\frac{r^2}{l^2}\bigg)dt^2+\bigg(1-\frac{r^2}{l^2}\bigg)^{-1}dr^2+r^2d\Omega_2,\qquad t_{\rm E}\rightarrow t_{\rm E}+ \beta\, ,
	\end{equation}
	with $\beta=2\pi l=1/T_\textup{dS}$, $d\Omega_2$ the line element of a 2-sphere and $l\sim 1/\sqrt{\Lambda}$. A coordinate transformation $r=l \cos\rho$ then allows us to rewrite the Euclidean dS space as the metric of a 4-sphere:
	\begin{equation}
		\label{mee}
		ds_{\rm E}^2=\sin^2\rho dt_{\rm E}^2+l^2d\rho^2+l^2\cos^2\rho d\Omega_{2}\, .
	\end{equation}
	To accommodate a conical singularity at $r=0$, the value of $\beta$ must be adjusted in order to ensure that the metric remains regular (non-singular) at the boundary $r=l$. The periodicity condition on $t_{\rm E}$ is replaced with
	\begin{equation}
		\label{repll}
		t_{\rm E}\rightarrow t_{\rm E}+\frac{2\pi l}{N} \, ,
	\end{equation}
	corresponding to the $\mathbb{Z}_{N}$ symmetry shift. In the limit of $N\rightarrow \infty$, the periodicity converges to $0$. The replacement $\beta \rightarrow \beta/N$, where $Z=e^{-\beta F}$ is the partition function and $F=E- S/\beta$ is the free energy, implies that the partition function of each cone corresponds to a $1/N$-power of the dS one:
	\begin{equation}
		\label{contr}
		Z_\textup{dS}\simeq (Z_\text{cone})^N,\qquad S_\textup{EdS}\simeq S_{S_4/\mathbb{Z}_N}\simeq N S_\text{cone}\, ,
	\end{equation}
	where $S_\text{cone}\sim 1$. Therefore, the dS spacetime and entropy emerge as a superposition of a large number of gravitational conic instantons. Let us note that these considerations are actually sustained in the lower-dimensional space $\text{dS}_{3}$, where the correspondence among conic instantons and the dS spacetime entropy was proven with exact computations and confirmed with the dual two-dimensional conformal field theory (CFT$_2$) \cite{Banados:1998tb}. Moreover, this conclusion is believed to be in agreement also with wave functions that promote the Wald entropy and the opening angle to quantum operators \cite{Brustein:2012sa}. Following this consideration, the number of hairons scales as the number of singular points of the orbifold, that is
	\begin{equation*}
		N\, \text{hairons} \leftrightarrow N\, \text{conic singularities} \leftrightarrow S_{4}/\mathbb{Z}_N\,.
	\end{equation*}
	
	
	As is known, modulus fields are Nambu--Goldstone bosons. Hence, it is considered to be natural in this picture to identify these moduli with hairons $\varphi_h^i$, with $i=1,\dots,N$. It is crucial to notice that the $N$ hairons are distinguishable as flavored under $\mathbb{Z}_N$. This aspect also gives the correct entropy count, which may otherwise be affected by factorial permutation factors. In the picture under consideration here, quantum qubits contained in the de Sitter spacetime correspond to quantum hairs and hairons/moduli. The mass of hairons is determined by the instanton curvature, which in the case considered is $m\sim \sqrt\Lambda$. Theoretically, the mass spectrum of metric fluctuations (which include the moduli) around an instanton background is determined by the Lichnerowicz operator. The eigenvalues of this operator on an orbifold sphere are proportional to the inverse of the Hubble radius. The mutual interactions of the fluctuations is, in a semi-classical or instanton gas approximation, negligible, and therefore those fluctuations may form a coherent state with characteristic wavelength $\lambda \sim 1/\sqrt\Lambda$. Actually, there may exist contributions from orbifold instantons with arbitrary large masses. However, these correspond to moduli with a relatively short wavelength, which do not contribute to cosmological infrared dynamics. In other words, the number of degrees of freedom scales holographically, as expected.
	
	Hairons, as gravitational modulus scalars, are typically described via sigma models, with the kinetic term of a Nambu--Goldstone boson scalar field. They have a proper energy-momentum tensor $T_{\mu\nu}(\varphi_1,\dots,\varphi_N)$, and therefore are coupled to the gravitational field. Thus, hairons can interact with SM matter through gravitational interactions in $\sqrt{-g}g_{\mu\nu}T^{\mu\nu}$ terms. In general, hairons may also have a Planck suppressed non-minimal coupling with the Ricci scalar, such as
	\begin{equation}
		\label{scal}
		{L}\sim\sqrt{-g}\varphi_i\varphi^i R + \dots\, ;
	\end{equation}
	the newly introduced term $\varphi_i\varphi^i/M_\textup{P}^{2}$ has a Planck mass suppression which is elided by the Newtonian constant of the Einstein--Hilbert Lagrangian.
	
\begin{center}
		\begin{figure}[h]
		\includegraphics[width=0.8\textwidth]{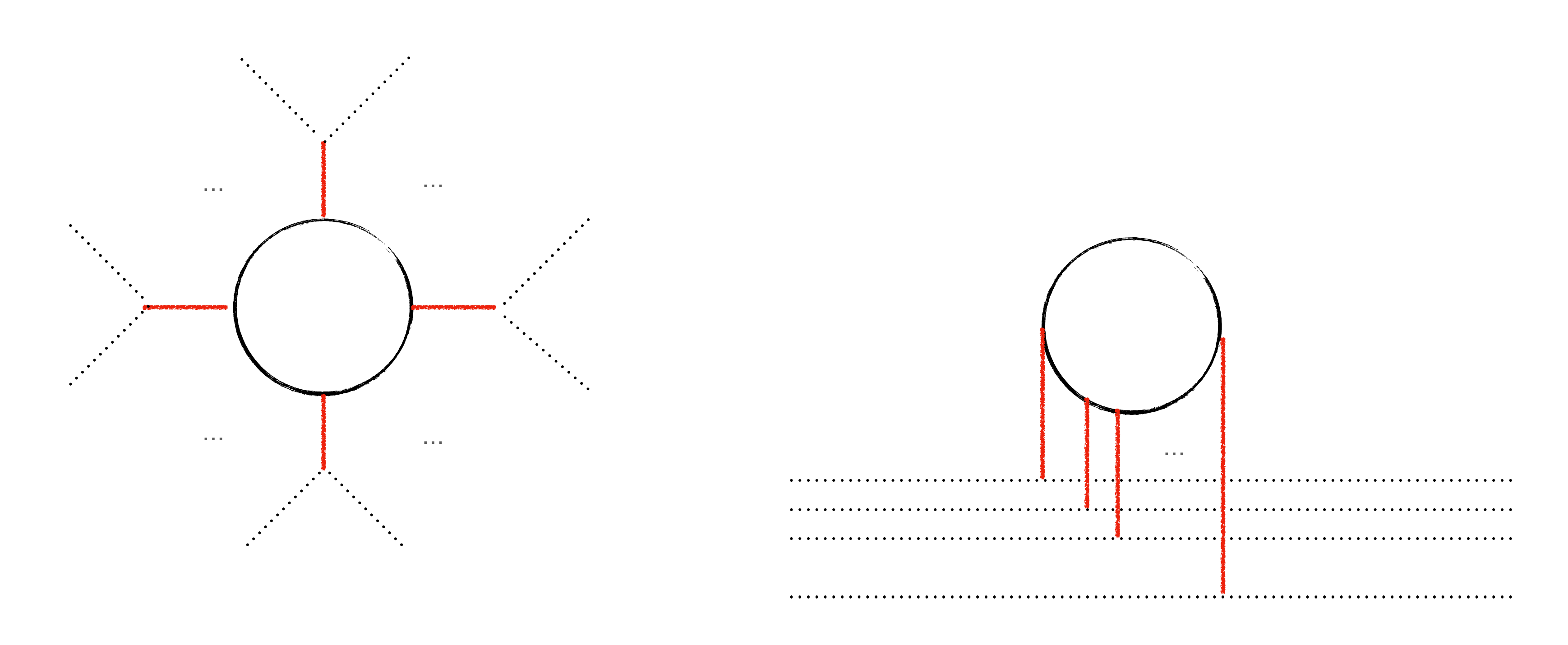}
		\caption{\textbf{Left}: insertion of a large number of gravitons (red lines) coupled to hairons (dotted lines), illustrating how quantum fluctuations interact with the horizon degrees of freedom. \textbf{Right}: a quantum loop diagram interacting with the hairon condensate, showing how vacuum energy contributions are thermalized and dissipated by the horizon's informational degrees of freedom, protecting the small value of the cosmological constant.}
		\label{fig:holographic_naturalness}
		\end{figure}
	\end{center}

	Let us now discuss why the couplings with the gravitational and matter sectors in Equation \eqref{scal} are fundamental for the holographic naturalness mechanism. The unnatural radiative loops receive the insertion of $N$ hairons through $N/2$ gravitons. As mentioned above in this section, $\mathcal{HN}$ starts with noticing that the evaluation of bubble diagrams in the SM in the true large-$N$ vacuum state is exponentially suppressed. Nevertheless, there is a diagram which in the large-$N$ state may dominate over all radiative corrections. This diagram can be built considering the insertion of $N/2$ gravitons within the bubble propagator line of any SM field (see Figure \ref{fig:holographic_naturalness}). Moreover, Equation \eqref{scal} implies that these gravitons can be non-minimally coupled with hairons, which are in turn in a cosmological condensate (actually, gravitons are also coupled to hairons as standard matter, but in that case the same diagram receives a significant suppression from the extra gravitational coupling of each insertion). The average energy of each hairon is then $\langle E \rangle\simeq m_h \sim \sqrt\Lambda \sim M_\textup{P}/\sqrt{N}$. Taking into account the presence of the background condensate, the $N$-insertion diagram is \emph{not} exponentially suppressed, since the number of couplings coming from graviton insertions is compensated by a stimulated emission factor (which is typically linear with respect to the number of each insertion):
	\begin{equation}
		\label{kkkaa}
		\alpha(\Lambda)^{N/2} {E}_{N/2} \rightarrow (N/2)^{-N/2} (N/2)^{N/2} \sim 1\, ,
	\end{equation}
	\textls[-15]{where ${E}_{N/2}$ is the stimulated emission enhancement stemming from the presence of a background condensate of hairons. This does not imply that such a diagram has, again, UV divergences, since now it is cut off by the effect of the average temperature of the condensate:}
	\begin{equation}
		\label{averagg}
		\Delta \rho_\text{vac}= (n_B - n_F) T_\textup{dS}^4 \sim(n_B-n_F)\Lambda^2\, ,
	\end{equation}
	where $n_{B,F}$ is the number of bosons or fermions. These corrections are natural, in the technical sense, because in the  limit of bare coupling, $\Lambda\rightarrow 0$, the corrections vanish. Therefore, when an SM particle is localized on the dS horizon, the particle interacts with this ``atmosphere'' of hairons. The Feynman diagrams with hairon loops represent how the particle's properties (like its contribution to the vacuum energy) are ``thermalized'' or ``dissipated'' into the horizon's degrees of freedom, preventing a catastrophic feedback loop.
	\vspace{-18pt}
	
	As a consequence of this process of stimulated emission, matter and radiation in the universe can also spontaneously emit soft (low-momentum) gravitons, adding more hairons to the background hairon condensate which constitutes dark energy. This phenomenon leads to an amplification of the emission probability, which scales as $N$ for every emission. That is, $N$ dynamically increases with the cosmological time, corresponding to a dynamical decrease in $\Lambda$ and a dynamical increase in entropy:
	\begin{equation}
		\label{increase}
		|N\rangle \rightarrow |N+1\rangle, \qquad\frac{\Delta \Lambda}{M_\textup{P}^2}\sim \frac{1}{N+1}-\frac{1}{N}<0,\qquad\Delta S=1 \, . 
	\end{equation}
	The process \eqref{increase} has the following asymptotic behavior: $N\rightarrow \infty$, $\Lambda \rightarrow 0$, $S\rightarrow \infty$. Indeed, this is an emission instability, predicting that a dS state with constant CC cannot exist in this model if not as a secularly growing steady approximation in time. This picture also suggests that if one starts with a small amount of hairons $N\sim 1$--$10$ corresponding to a considerably large CC, its value spontaneously decreases after a certain relaxation time. However, in the case of W-gravity one can start directly from a small enough value of $\Lambda$ obtained from a see-saw mechanism, with the $\mathcal{HN}$ paradigm guaranteeing its stability against quantum effects. Note that in the case of $N\sim 10^{120}$, as in the case of the CC, the time scale for a first transition is of the order $\tau \sim N/\sqrt\Lambda$, i.e., much larger than the age of the universe. However, if one considers a transition $|1\rangle \rightarrow |N\rangle$, the corresponding change $|\Delta \Lambda|= |1/N -1| M_\textup{P}^{2} \sim M_\textup{P}^2$ suggests that such a transition is almost instantaneous, taking only a Planckian time to happen. While this result can be considered as another way to solve the CC problem through dynamical relaxation, one has to be cautious and ponder the meaning of a stimulated emission starting from a universe with only few initial qubits.
	
	One may conceive the dS spacetime starting with an exceptionally large value of the CC and understand the relaxation process as exponentially efficient in cosmological time. Then, naturally, in a pre-geometric scenario, this dynamic is considered to start only after the fundamental SSB. This perspective can solve one of the major theoretical problems of the PGG approach: why such a large value of the gravitational Higgs VEV $v$ is to be chosen? One unsatisfactory answer may come from invoking the anthropic principle, which is often seen as a last resort. The pre-geometric phase transition may produce a multiverse of domains with different values of $v$. Domains with a relatively small $v$ (large $\Lambda$) have insufficient entropy and an exceptionally large value of the CC to form complex structures or intelligent observers. Nevertheless, the stimulated emission mechanism provides a way to completely avoid engaging in anthropic arguments. In general, one can imagine the universe having been originated with an arbitrary value of $v$, which in general, can also be as small  as $v^n \ll10^{120}$ (with $n=1,3$ for MM- or W-gravity respectively). This then corresponds to less qubits of information and less nonzero entropy, as well as a much larger Hubble rate than the currently observed one. As $N$ increases because of the mechanism of stimulated emission, $\Lambda$ decreases and $v$ increases. This picture may be dubbed walking gravitational vacuum. Though one has to be cautious in extending these considerations for regimes with $v\sim 1$ and Planckian CC, these regimes can be safely considered in the case $1\ll v\ll 10^{120}$. Following this argument, we conclude that the VEV of the pre-geometric field $\phi^A$ may possibly never be stable, deforming its sombrero potential in secularly growing minimal dynamics. We are then tempted to suggest that this may also naturally solve the problem of the arrow of time: after the SSB, the spacetime metric is generated and the concept of clocks can be defined; in addition, the stimulated emission is an irreversible process since the probability of emitting $N$ hairons to enrich the condensate is $e^N$ higher than the re-absorption process. Therefore, the increase in entropy from stimulated emission, which is related to the dynamical increase in the VEV, may naturally provide the physical foundation behind the common perception of the arrow of time.

	\section{Hamiltonian Analysis, Hairons and Emergent Degrees of Freedom}\label{sec5}
	In this Section, we proceed to remark the Hamiltonian structure of theories of PGG, these theories' degrees of freedom and the consistency with the existence of the hairon condensate. The Lagrangian density of pre-geometric theories (see Section \ref{sec3}) exhibits a degenerate structure, as it is linear in the temporal derivatives (velocities) of the pre-geometric fields. To this end, for carrying out the Hamiltonian analysis it is necessary to employ Dirac's systematic procedure for constrained Hamiltonian systems \cite{dirac:hamiltonian}.
	
	For the W model, the complete Hamiltonian density \cite{Addazi:2025vbw} takes the form
		\begin{equation}\label{Hdens} 
			{H} = -A_0^{AB}\Big[\partial_i\Pi^i_{AB}(\phi,A) + 2\Pi^i_{BC}(\phi,A)A_{Ai}^C + \eta_{BC}\Pi_A(\phi,A)\phi^C\Big] + \lambda^A Z_A + \lambda_i^{AB} Z^i_{AB} + \lambda_0^{AB} Z^0_{AB},
	\end{equation}
	where $\lambda^A$, $\lambda_i^{AB}$ and $\lambda_0^{AB}$ represent Lagrange multipliers and $\partial_i\equiv \partial /\partial x_i$. The conjugate momenta $\Pi_A$ to $\phi^A$ and $\Pi^\lambda_{AB}$ to $A_\lambda^{AB}$, evaluated on the constraint surface, are defined as
	\vspace{-5pt}
	\begin{equation}
		\Pi_A(\phi,A)\equiv 2\epsilon_{ABCDE}\epsilon^{0ijk}\nabla_k \phi^D\phi^E[k_\textup{W}F_{ij}^{BC}-2\text{sgn}(J)k_\textup{SB}v^{-4}\nabla_i\phi^B\nabla_j\phi^C(\phi^2\mp v^2)^2],
	\end{equation}
	and
	\begin{equation}
		\Pi^\lambda_{AB}(\phi,A)\equiv2k_\textup{W}\epsilon_{ABCDE}\epsilon^{0\lambda jk}\nabla_j\phi^C\nabla_k\phi^D\phi^E,
	\end{equation}
	respectively.
	
	Specifically, the spatial and temporal components of the latter are given by
	\begin{equation}
		\Pi^i_{AB}(\phi,A)=2k_\textup{W}\epsilon_{ABCDE}\epsilon^{0ijk}\nabla_j\phi^C\nabla_k\phi^D\phi^E,\qquad\Pi^0_{AB}(\phi,A)=0.
	\end{equation}
	Dirac's procedure then leads to three primary constraints:
	\begin{equation}
		Z_A\equiv\Pi_A-\Pi_A(\phi,A)\approx0,\qquad Z^i_{AB}\equiv\Pi^i_{AB}-\Pi^i_{AB}(\phi,A)\approx0,\qquad Z^0_{AB}\equiv\Pi^0_{AB}\approx0,
	\end{equation}
	where ``$\approx$'' denotes a `weak' equality on the constraint surface.
	
	The preservation of the primary constraint $Z^0_{AB}$ under time evolution generates a secondary constraint:
	\begin{equation}
		\dot{Z}^0_{AB} = \{Z^0_{AB}, H\}= \partial_i\Pi^i_{AB}(\phi,A) + 2\Pi^i_{BC}(\phi,A)A_{Ai}^C + \eta_{BC}\Pi_A(\phi,A)\phi^C \approx 0,
	\end{equation}
	where the dot denotes the time derivative and ``$\partial_i$'' the spatial ones. The total Hamiltonian density \eqref{Hdens} then takes on a simplified form:
	\begin{equation}\label{eq:total-H}
		\begin{split}
			{H} &= -A_0^{AB}\dot{Z}^0_{AB} + \lambda^A Z_A + \lambda_i^{AB} Z^i_{AB} + \lambda_0^{AB} Z^0_{AB} + \tilde{\lambda}_0^{AB}\dot{Z}^0_{AB} \\
			&\equiv \lambda^A Z_A + \lambda_i^{AB} Z^i_{AB} + \lambda_0^{AB} Z^0_{AB} + \tilde{\lambda}_0^{AB}\dot{Z}^0_{AB},
		\end{split}
	\end{equation}
	where terms proportional to $\dot{Z}^0_{AB}$ are re-absorbed through a redefinition of the Lagrange multiplier $\tilde{\lambda}_0^{AB}$. The field $A_0^{AB}$ no longer appears explicitly in the expression \eqref{eq:total-H}: this reveals its status as an exclusively gauge degree of freedom, and hence the Lagrange multiplier $\lambda_0^{AB}$ remains arbitrary.
	
	The phase space of the theory is described by ninety dynamical variables \cite{Addazi:2025vbw}, namely the fields $(A_0^{AB}, A_i^{AB}, \phi^A)$ and their corresponding conjugate momenta $(\Pi_{AB}^0, \Pi_{AB}^i, \Pi_A)$. Gauge freedom is characterized by twenty gauge-fixing conditions, which eliminate the unphysical degrees of freedom associated with $A_0^{AB}$ and $\Pi_{AB}^0$. The constraint structure consists of ten first-class constraints ($Z_{AB}^0$), which generate gauge transformations, and forty-four second-class constraints (combinations of $Z_{AB}^i$, $Z_A$ and $\dot{Z}_{AB}^0$). The number of physical degrees of freedom is thus given by
	\begin{equation}
		2 \times N_{\textup{dof}} = N_{\textup{dyn}} - N_{\textup{gauge}} - 2N_{\textup{1st}} - N_{\textup{2nd}} = 90 - 20 - 2\times10 - 44 = 6,
	\end{equation}
	yielding $N_{\textup{dof}} = 3$. This count corresponds to a massless spin-2 graviton ($2$ degrees of freedom) and a massive scalar field $\rho$ ($1$ degree of freedom), in an analogy to scalar-tensor theories of gravity. This result is background-independent and valid for the MM model too.
	
	After the SSB, the pre-geometric Hamiltonians reduce to the Einstein--Hilbert form of the ADM formalism, with the addition of the extra scalar field $\rho$ whose mass is expected to be close to the Planck scale \cite{Addazi:2025vbw}. This result establishes a compatibility with the loop quantum gravity framework too, in particular through the natural emergence of Ashtekar's electric field variables from the components of $\Pi^i_{A0}$ post-SSB.
	
	The Hamiltonian analysis in this Section is crucial and unambiguous: pre-geometric theories contain three physical degrees of freedom, that is the two polarizations of the graviton and a massive scalar mode $\rho$. This may initially appear to be at odds with the prior discussion of the hairon field $\varphi_h$ as a new entity. However, this is not a contradiction but a refinement of its nature. Actually, the hairon field does not constitute a new, independent degree of freedom. It is an emergent composite quasi-particle that manifests itself after the SSB as a specific collective excitation of the pre-geometric condensate. Indeed, hairons are Nambu--Goldstone boson modulus fields that characterize the orbifold gravitational instantons---and these are to be seen as semi-classical quantum effects, rather than fundamental building blocks of the theoretical framework under examination.
	
	A tempting but ultimately misleading analogy may be considering $\varphi_h$ as a bound state of $\rho$ and $A_\mu^{AB}$ quanta, akin to the proton in quantum chromodynamics. This picture, actually, fails to explain the hairons' light mass, as the mass of such a bound state is expected to be at least of the order of the mass of its heaviest constituent, that is $m_\rho \sim v \sim M_{\textup{P}} \gg \sqrt{\Lambda}$. Instead, a correct analogy is drawn not from high-energy physics but from condensed matter physics. In a crystal lattice, the fundamental constituents are atoms with a comparably large mass. Yet the collective vibrational modes of the lattice---called phonons---are massless (acoustic phonons) or have an exceptionally small mass gap (optical phonons), with energies determined by the lattice spacing and bond strengths, not by the atomic mass scale. Similarly, in a ferromagnet, the fundamental electron mass is quite large, but the relatively low-energy spin-wave excitations---called magnons---have a dispersion relation $\omega \sim D k^2$ governed by a comparably small stiffness constant $D$. The analogy with magnets and spin waves can further elucidate the relationship between pre-geometric fields and hairons. $\phi^A$ is akin to the magnetization field $M(x)$ of a ferromagnet. Before the phase transition, its spins are disordered ($M = 0$). Below the Curie temperature, the spins can spontaneously align, leading the field $M$ to acquire a nonzero VEV $\langle M \rangle$ that breaks the rotational symmetry of the system. This value is a macroscopic, static order parameter. In this sense, $\varphi_h(x)$ is then similar to a spin wave (a magnon) in the ferromagnet. A magnon, actually, is a quantum excitation of the condensed ground state defined by $\langle M \rangle$---it is a `ripple' in the order parameter. One cannot have magnons without the underlying magnetic order, because they are perturbations thereof. The condensation of $\phi^A$ is thus what sets up the ``geometric order'' of the universe as $\langle \phi^4\rangle=v$. A hairon $\varphi_h(x)$ is an excitation or a fluctuation of this order parameter on the dS horizon. Indeed, pseudo-particles in condensed matter systems are Nambu--Goldstone bosons just like the hairons. The key distinction lies in their origin: in condensed matter physics, Nambu--Goldstone bosons arise from symmetries that are spontaneously broken by the crystalline structure of the material itself. Hairons, conversely, originate from symmetries broken by instantons together with the Higgs phase of spacetime induced by $\langle \phi^4 \rangle=v$. Consequently, hairons only emerge at a semi-classical level alongside those instantons, whereas condensed matter systems and their Nambu--Goldstone bosons are already present at the classical level.
	
	The key to understanding the hairons' comparably light mass lies in the structure of the effective action of the theory. After the SSB, a term $\Lambda \varphi_h^2$ emerges not from the fundamental mass of $\rho$ but from the interplay between the comparably large VEV $v$ and the relatively small pre-geometric coupling constant $k_\textup{W}$ (or $k_\textup{MM}$, for MM-gravity). This is why the resulting mass scale is eventually set by the CC:
	\begin{equation*}
		m_h^2 \sim \Lambda = -\frac{3M_\textup{P}^2}{4k_\textup{W} v^3}.
	\end{equation*}
	The largeness of $v$ is precisely compensated by the smallness of the effective coupling $1/(|k_\textup{W}| v^3) \sim 1/S_\textup{dS}$, leading to a hierarchically small mass $m_h \sim \sqrt\Lambda$. The hairon mass is therefore a derived, emergent property of the condensed vacuum state, not a parameter of the fundamental Lagrangian of PGG. It is light because it is the pseudo-Nambu--Goldstone boson of symmetries which are broken by orbifold instantons, and its value is protected by the exceptionally high entropy $S_\textup{dS}$ of the dS vacuum.
	
	\section{Remarks on $\mathcal{HN}$ Phenomenology}\label{sec6}
	$\mathcal{HN}$ posits that a hidden sector simulates information loss for an observer confined to the SM sector. From this limited perspective, phenomena such as non-conservation of probability, violation of unitarity and CPT, and apparent energy non-conservation may emerge. However, these phenomena are only effective descriptions; the complete theory remains fully unitary and consistent.
	
	Consequently, $\mathcal{HN}$ directs the search for new physics toward signatures of apparent quantum decoherence within SM observables. This approach is fundamentally different from paradigms like the alleged TeV-scale supersymmetry or composite-Higgs models, which are primarily probed by high-energy colliders like the LHC. Instead, $\mathcal{HN}$ signatures are best sought in high-precision, relatively low-energy experiments. The following lists some of the most promising possibilities for testing $\mathcal{HN}$.
	\begin{itemize}
		\item Neutral meson oscillations, such as kaon--antikaon transitions \cite{E,Mavromatos:2018rds}.
		\item Baryon number violation: searches for neutron--antineutron oscillations \cite{Babu:2016rwa,Addazi:2015oba,Addazi:2022dfj}.
		\item Quantum interferometry: using high-precision interferometers \cite{Verlinde:2019xfb} or entangled systems \cite{E1,E2}.
		\item Astroparticle physics: anomalies in velocity dispersions of very-high-energy (above TeV) cosmic neutrinos \cite{Addazi:2021xuf,AlvesBatista:2023wqm}.
	\end{itemize}
	
	Regarding quantum interferometry, it is crucial to distinguish the $\mathcal{HN}$ framework considered here from models of intrinsic Planck-scale `holographic noise' \cite{Ng:2003,Hogan:2009mm,Goklu:2009jz}, which predict universal decoherence and metric fluctuations. In contrast, the $\mathcal{HN}$ mechanism posits that the holographic degrees of freedom (the hairons) are emergent, infrared modes in a coherent state, thus evading the stringent experimental bounds on universal spacetime granularity. The models of holographic noise (like those proposed by Y.\ Jack Ng \cite{Ng:2003}, Craig Hogan \cite{Hogan:2009mm} and Ertan G\"{o}kl\"{u} and Claus L\"{a}mmerzahl \cite{Goklu:2009jz}) posit intrinsic spacetime fluctuations at quite short scales, often modeled as stochastic metric perturbations. A suite of universal effects, such as transverse jitter in interferometer arms, modified wave packet spreading, apparent violation of the equivalence principle, and decoherence in matter-wave interferometry have been predicted. A key tenet is that spacetime is fundamentally `foamy', leading to observable noise even in the absence of an emergent geometry. These models often entail a violation of unitarity or CPT symmetry and typically introduce fundamental uncertainty cells, whose scale, in extreme cases, is about $10^{-16}$\,m---a scale already ruled out by experiments. In the $\mathcal{HN}$ and PGG frameworks under consideration, granularity is not a fundamental property of spacetime, but emerges from the degrees of freedom on the horizon (the hairons). This granularity is thus informational, not geometric. Consequently, we do not postulate the existence of fundamental cells of size of about $(10^{-16}\,\text{m})^2$, but rather emergent cells where information is packed within a Planck area of about $(10^{-35}\,\text{m})^2$. The Planck length itself, actually, is not fundamental but only emergent after the SSB. The hairon condensate is a coherent state with a cosmological-scale wavelength and a very light mass of about $10^{-33}$\,eV. The behavior of the condensate is not thermal but coherent, signifying that effects like `random walk noise' are inapplicable. This allows the model considered here to evade current experimental limits constraining stochastic models of holographic noise.
	
	Furthermore, if the hidden sector states---the hairons---carry angular momentum, they may mediate apparent Pauli-violating transitions too \cite{Addazi:2017bbg,Napolitano:2023lar}. For instance, hairon vortices in vacuo allow for a scattering SM particle to exchange a small amount of angular momentum, leading to observable effects in underground experiments.

		\section{Conclusions}\label{sec7}
		In this paper, we presented a unified framework that addresses the cosmological constant problem by weaving together the principles of holography, quantum information and PGG. The synthesis leads to several profound conclusions and opens up new avenues for research.
		
		\vspace{0.1cm}
		
		\textbf{1. A synergistic solution.} The two most vexing aspects of the CC problem---its initial smallness and its stability  against quantum corrections---are solved by two interconnected mechanisms within a single, consistent framework. Pre-geometric gravity provides an explanation for the smallness of the CC via a see-saw mechanism: a comparably large VEV $v$ of the pre-geometric Higgs field $\phi^A$ naturally leads to a quite small cosmological constant $\Lambda \sim 1/v$ (or $1/v^3$). Holographic naturalness explains the stability of the CC: the same quite large VEV $v$ is identified with the de Sitter entropy $S_\textup{dS}$, and the exponential suppression $\exp(-S_\textup{dS})$ of decay amplitudes protects the relatively small value of the CC. The universe maintains a quite small CC because it has a large enough information capacity and, also, is stable enough for the same reason.
		
		\vspace{0.1cm}
		
		\textbf{2. The nature of hairons.} Hairons are not fundamental degrees of freedom of the $\mathcal{HN}$+PGG framework. Instead, hairons emerge as pseudo-Nambu--Goldstone bosons, which are bosonic modes associated with orbifold gravitational instantons. The dynamics of hairons is governed by a sigma-model Lagrangian and their mass is naturally suppressed, being inversely proportional to the Hubble radius. These particles couple to both SM fields and gravity, and under certain conditions may form a Bose--Einstein condensate with a coherent wavelength on the cosmological scale of the Hubble radius. The  condensate of hairons does not produce a static dark energy; instead, quantum radiative processes introduce instabilities that generate a dynamical form of dark energy.
		
		\vspace{0.1cm}
		
		\textbf{3. Emergence of spacetime and information.} The results obtained here strengthen the radical view that spacetime is an emergent phenomenon. The pre-geometric phase, described by a gauge theory without a metric, undergoes a phase transition to a geometric, Higgs phase. This transition simultaneously generates the Einstein--Hilbert action, the Planck scale, the cosmological constant and the information content (entropy) of spacetime. The Higgs field $\phi^A$ is thus truly an `information field', whose condensate defines both the geometry and its associated quantum information. The holographic principle, which posits an entropy-area law, is interpreted in PGG as an emergent phenomenon rather than a fundamental postulate. This emergence is theorized to result from a process of SSB, from which the de Sitter entropy-area law itself originates. Indeed, also $\mathcal{HN}$ and hairons do not exist in the  pre-geometric unbroken phase. The deep connection between entropy and geometry points at the existence of novel principles to be pursued in the quest for a theory of quantum gravity, one in which the laws of physics are ultimately laws of information processing and stability.
		
		\vspace{0.1cm}
		
		In conclusion, by identifying the pre-geometric Higgs VEV as the source of the holographic entropy of spacetime, we developed a compelling and self-consistent narrative for the origins and the stability of the CC. The problem is solved not by introducing fine-tunings or new symmetries in the particle spectrum, but rather by acknowledging the elemental role of quantum information in the very fabric of spacetime. A considerably small and stable cosmological constant is a natural consequence of a universe that is, at its core, a system of exceptionally large information content whose geometry is only a secondary, emergent property.

	\vspace{0.2cm}
	
	{\bf Acknowledgements}.
	During the preparation of this work, the authors used DeepSeekV3 to improve readability and language. After using this tool, the authors reviewed and edited the content as needed and take full responsibility for the content of the publication. The work by A.A.\ is supported by the National Science Foundation of China (NSFC) through grant No.\ 12350410358; the Talent Scientific Research Program of the College of Physics, Sichuan University, Grant No.\ 1082204112427; the Fostering Program in Disciplines Possessing Novel Features for Natural Science of Sichuan University, Grant No.\ 2020SCUNL209; and the 1000 Talent program of Sichuan province 2021. S.C.\ and G.M.\ acknowledge the support of Istituto Nazionale di Fisica Nucleare (INFN), Sezione di Napoli, Iniziative Specifiche QGSKY and MoonLight-2. This paper is based upon work from COST Action CA21136---Addressing observational tensions in cosmology with systematics and fundamental physics (CosmoVerse), supported by COST (European Cooperation in Science and Technology).

\end{document}